\documentstyle[prb,aps,manuscript]{revtex}

\begin{document}                

\title{ A Model for the Optical Absorption in Porous Silicon.  }

\author{Shouvik.Datta $^{*}$ and K.L.Narasimhan}
\address{Solid State Electronics Group .}
\address{Tata Institute of Fundamental Research, Homi Bhabha Road, Colaba,
Mumbai 400 005. INDIA .}

\maketitle
\begin{abstract}

\hskip 2cm  In this paper, we analyse the optical absorption in porous
silicon. This is the first attempt to explicitly demonstrate that it is
not possible to extract the band gap of such $low$ $dimensional$
nanostructures like porous silicon from a Tauc plot of
$\sqrt{\alpha\hbar\omega}$ vs $\hbar\omega$. So we  model the absorption
process assuming that porous silicon is a pseudo 1D material system having
a distribution of band gaps.  We show that in order to explain the
absorption we specifically need to invoke - (a) k is not conserved in
optical transitions, (b) the oscillator strength of these transitions
depends on the size of the nanostructure in which absorption takes place
and (c) the distribution of band gaps significantly influences the optical
absorption. A natural explanation of the temperature dependence of
absorption in porous silicon also follows from our model.

\end{abstract}

\vskip 0.5cm
\flushleft{ $^{*}$ e.mail :  shouvik@tifr.res.in  }
\vskip 0.5cm
\hskip 10.5cm  $Typeset$ $using$ $REVTEX$

\newpage

\section {Introduction}

\hskip 2cm  The discovery of efficient $photoluminescence$ $^{( 1, 2 )}$
from porous silicon (PS) has attracted the attention of many researchers.
The photoluminesecence peak occurs at an energy greater than the band gap
of c-silicon and can be tuned through the visible spectrum by changing the
preparation conditions. $Canham$ $ ^{( 1 )}$ proposed that PS is a
nanostructured material and the band gap of these nanostructures is
enhanced due to quantum size effects. He argued that this would account
for the fact that the luminescence energy is greater than the band gap of
c-silicon.  Evidence for the increase in band gap was first reported by
$Lehmann$ - from optical absorption measurements $^{( 3 )}$. Transmission
electron microscopy measurements reveal that  the diameter of the columnar
nanostructures that make up PS depends on preparation conditions and can
be as small as 2 - 4 nm $^{( 2, 3 )}$.  PS can hence be thought of as an
assembly of pseudo 1D quantum wires.  The progress in this area has been
the subject of many recent reviews $^{( 4 - 8 )}$.  In this paper we
confine ourselves to the understanding of the optical absorption process
in PS.\\

\par

\hskip 2cm  There have been many attempts to obtain the band gap of PS
from optical absorption measurements $^{( 9 - 13 )}$. In these
measurements, the authors assume that the absorption coefficient
($\alpha$) of PS satisfies the same relation as in 3D bulk c-silicon viz.

$$ \sqrt{\alpha\hbar\omega} = A(\hbar\omega -E_g)
\eqno(1)
$$

\hskip 2cm    There the Tauc plot of $\sqrt{\alpha\hbar\omega}$ vs
$\hbar\omega$ is a straight line and the intercept on the energy axis
gives the band gap $E_g$.  The use of this equation for a 1 dimensional
system like PS suffers from the following criticisms .\\
\par
\hskip 2cm     1) In general ,the absorption co-efficient depends on the
joint density of states (JDOS) of the material. Eq.1 is valid only when
the density of states g(E) varies $^{ ( 14 )}$ with energy near the band
edges as g(E) = M$\sqrt{ E }$, where E is measured from the band edge and
M is a constant. If we treat PS as a  1D material ( upto a first
approximation assuming parabolic band structure ) it would be more
$appropriate$ to write the density of state g(E) = $N\over\sqrt{E}$ ( N is
a constatnt ) than the common practice of using 3D density of state.
Clearly the JDOS in PS is expected to be different from that of bulk
c-silicon.  For a 1D system having a single direct band gap $E_g$, it is
easy to show that the JDOS is proportional to $N\over\sqrt{(E-E_g)}$.  In
such a case, a plot of $\alpha(E)$ vs E should peak at $E_g$. On the
otherhand, for an indirect band gap material ( assuming that the momentum
k is not conserved in this pseudo 1D system in an optical transition ), it
is easy to show that the JDOS is a constant and independent of incident
energy ( following reference (14) ) .  Therefore , in either case ,
$\sqrt{\alpha\hbar\omega}$ is not linearly related with $\hbar\omega$ at
all. Clearly the use of eq.  1 to extract the band gap is not justified
for PS
. \\
\par

\hskip 2cm    2) The nano structures that make up PS have a
$distribution^{( 2, 3, 8 )}$ of diameters (d). If the band gap is related
to quantum size effects, then it is clear that the energy upshift $\Delta
E$ ( which can be written as $\Delta$E = C/d$^{X}$ ) is different for each
of the nanostructures that make up PS.( The value of X in the simplest
approximation is 2 , further details of this are discussed in section III
). In PS, we are actually dealing with a heterogeneous system with a
distribution of band gaps. Hence it may not be meaningful to visualize PS
as having a single band gap and that this can be extracted from equation
1. \\
\par
\hskip 2cm At this junction, we would also like to point out that the above
objections against the common practice of using  $^{( 15 )}$ the eqn. 1
to get the band gap are equally vaild in case of other low dimensional
systems like quantum dots etc.  \\

\par

\hskip 2 cm   In an attempt to address these questions in case of PS, we
have investigated the optical absorption in PS specifically using the 1D
density of states . In this paper we clarify the absorption process in PS
assuming that PS has a distribution of band gaps. Section II contains the
experimental results which will be used for comparing our simulations.
Here we report a $non-destructive$ way of measuring the porosity of PS
thin films using transmission measurements and effective mass theory.
Section III outlines the model used for calculating optical absorption and
section IV the results of our simulations . On the basis of the
simulations we will further consolidate our viewpoint that the Tauc plot
can not be used to determine the band gap of such low dimensional systems .
We also show that we can easily explain the temperature dependence of the
absorption in PS using our model .Finally we summarise our conclusions in
section V .

\section{ Experimental Results}

\hskip 2cm  We have done some experiments to facilitates  the comparison
of our simulations with real experimental data . 

\hskip 2cm     PS is made by electrochemical etching of p type c-si in a
1:1 HF : Ethanol solution at a current density between 10 - 50
mA/cm$^{2}$.  Free standing samples are lifted off the substrate by
increasing the current density to $>$ 200 mA/cm$^{2}$.  The samples are
then washed in water to get rid of HF , rinsed in n-pentane and finally
dried in air on a glass substrate for transmission measurements .
\par

\hskip 2cm   Transmission measurements are carried out using a CVI 240
Digikrom monochromator with a tungsten lamp as a source, silicon detector
and SR 530 lock -in amplifier .For some samples we also use a Cary 1756
spectrometer. We have checked that internal multiple scattering does not
dominate the absorption in our samples. This is done by making two PS
samples of two different thickness ($5\mu$ and $10\mu$ respectively) and
ensuring that the transmission curves are the same for both samples at low
absorption $^{( 11 )}$ .

\par 

\hskip 2cm  The absorption coefficient $\alpha(\lambda)$ as a function of
wavelength $\lambda$ is obtained from the normalised transmittance
T($\lambda$) using $^{( 14 )}$

 $$T(\lambda) = {(1 - R)^2 exp(-\alpha (\lambda)(1 - P)t)\over 1 - R^2
exp(-2\alpha (\lambda)(1 - P)t)}
\eqno(2)
$$

\hskip 2cm  The reflectivity (R) is obtained in the low $\alpha$ region
using the relation R = (1-T)/(1+T).  P is the porosity of the layer and t
is its thickness.\\

\par
\hskip 2cm     The porosity of PS is usually determined by
gravimetric measurements which also destroy the sample.  We now show that
optical transmission experiment allows us to measure porosity in a
$nondestructive$ fashion .\\
\par
\hskip 2cm     The normalized transmittance ( in eq.2) for $\alpha$ = 0
reduces to

$$ T_L = {(1 - R)\over (1 + R)} \ or\ R = {1 - T_L\over 1 + T_L}
\eqno(3)
$$

\par
\hskip 2cm    If n is the real part of the refractive index , then we can
also write for the reflectivity ( for $\alpha$ = 0 ) as

$$ R =  {(n-1)^2\over(n+1)^2}
\eqno(4)
$$

\hskip 2cm   It follows from eq.4 that the refractive index (n$_{PS}$)
of PS is

$$ n_{PS} = {1 + \sqrt{R}\over 1 - \sqrt{R}}
\eqno(5)
$$
\par

\hskip 2cm     We model PS as consisting of two media - air ( having a
refractive index ( $n_{air}$ = 1 )) and crystalline silicon ( n$_{Si}$ =
3.44 ) . It was shown earlier $^{( 13 )}$ that in the framework of effective 
medium approximation one can write ,

$$
(\epsilon_{PS}-1)/(3\epsilon_{PS})=(1-P)[(\epsilon_{Si}-1)/
(2\epsilon_{PS}+\epsilon_{Si})]
\eqno(6)
$$

\noindent where P is the porosity and the dielectric constant $\epsilon= n^{2}$
 Therefore,

$$ P = 1-[(2\epsilon_{PS}+\epsilon_{Si})(\epsilon_{PS}-1)/(\epsilon_{Si}-1)
(3\epsilon_{PS})]
\eqno(7)
$$

\hskip 2cm  The porosity can now be obtained using equations 4 to 7.
\par

\hskip 2cm    To check the validity of eq.7, we calculate the porosity (
using eq.7 ) and compare it with the corresponding values of P obtained
gravimetrically.  Figure 1 is a plot of the porosity determined from the
transmission curves and the corresponding experimentally determined
porosity by gravimetric measurements as reported in the literature$^{( 9,
16 )}$ .  We see from figure 1 that the points nearly fit a straight line
with slope one .  We hence conclude that optical transmission measurements
can reliably be used to obtain the porosity of porous silicon in a
$non-destructive$ fashion .
\par

\par
\hskip 2cm    Figure 2 is a plot of log($\alpha$) vs E at 300 K and at 100
K.  The absorption at low temperature is reduced with respect to that at
room temperature.  We see that the two curves exhibit a rigid shift in
agreement with other results $^{(10)}$. An explanation of this phenomena
on the basis of our simulated results is given in section IV  .
\\
\par
\hskip 2cm   We now attempt to understand the optical absorption process
in PS with the help of some model calculations and simulations, assuming
that these 1D materials have a distribution of band gaps .\\

\section{ A Model for Absorption in PS having a distribution of band gaps}

\hskip 2cm  We model the nanostructures that make up PS to be  1D
parallelopipeds of square cross sections of side d and constant length L.
Since d is typically  20 - 40 $\AA$, the band gap of the material is
enhanced due to quantum size effects .  We assume that the
effective band gap ( lowest energy gap ) for a particular parallelopiped
is the minimum separation between the conduction band and the valence band
states for quantum numbers n$_{x}$ = 1 and n$_{y}$ = 1. We also assume
that the absorption takes place in the bulk.
\par
\hskip 2cm  The absorption coefficient $\alpha$(E) for a particular value
of incident energy E can be written as

$$
\alpha (E) = \int_{ E_G<E} \rho^J (E - E_G) F(E_G) P(E_G)dE_G 
\eqno(8)
$$
\par
\noindent  where the integration is done over all the band gaps (E$_{G}$)
of the system below the incident energy E. In equation 8, $\rho_{J}$ is
the joint density of states, F(E$_{G}$) the oscillator strength for the
optical transition and P(E$_{G}$) the distribution of band gaps E$_{G}$ in
the material.  We now discuss eq.8 in some detail.
\par
\hskip 2cm    We have already pointed out in section I, if g(E) =B/$\sqrt
E$ then the JDOS is $\rho^J =A (E - E_G)^{-W}$ where A is a constant.The
value of W depends on whether PS is a direct or an indirect band gap
material. For direct band gap material W = 0.5 and for indirect band gap
material W = 0
.
\par
\hskip 2cm    F(E$_{G}$) is the oscillator strength governing the
absorption.  For indirect band gap material, the oscillator strength $^{(
17 )}$ can be enhanced for small d . This happens because the overlap
between the electron and hole wavefunctions in k space increases as a
result of quantum confinement and contributes to an increase in the
oscillator strength at small d (large $E_{G}$)  and can be written $^{(
18, 19 )}$ as F = f/d$^{\gamma}$ where $\gamma$ = 5 to 6.  This manifests
itself as the dominant no phonon line which has been seen recently in the
luminescence spectra of PS $^{( 20 )}$.
\par
\hskip 2cm   The third term in eq.8 viz. $P(E_G)$, is the distribution of
band gaps in PS. The band gap distribution is obtained by assuming a
distribution of sizes for d and a relation governing the upshift in energy
$\Delta$ E with the size d ( due to quantum confinement ).
\\

\par
\hskip 2cm   In our model we have considered two possible distribution of
sizes. One is Gaussian and the other is Lognormal$^{( 21 )}$ distribution
given by $P^G$(d) and $P^L$(d) respectively.

$$ P^G(d) = {1\over (\sqrt{2\Pi\sigma})} exp(-{(d - d_0)^2\over
2\sigma^2})
\eqno(9)
$$

\noindent where d$_{0}$ is the mean size and the $\sigma$ the
standard deviation.

$$ P^L(d) = {1\over\sigma_Ld\sqrt{2\Pi}} exp[-{(ln(d)-m_0)^2\over
2{\sigma_L}^2}]
\eqno(10)
$$

\noindent where m$_{0}$ = ln(d$_{0}$), and $\sigma_{L}$ = ln($\sigma$),
$d_{0}$ being the mean size and $\sigma$ the standard deviation.\\
\par
\hskip 2cm   Electron microscopy measurements$^{( 2, 3, 22 )}$ suggest
that for p PS, d$_{0}$ is around 30 $\AA$ and $\sigma \approx$ 4 $\AA$.
Figure 3 shows the distribution of sizes for  Lognormal and Gaussian
distributions for d$_{0}$ = 30 \AA ~and $\sigma$ = 4 \AA .

\par
\hskip 2cm   The energy up shift for the confinement $\Delta$E can be
written as

$$
\Delta E = E_G - E_g = C/d^X
\eqno(11)
$$

\noindent where E$_{g}$ is the crystalline Si fundamental
indirect band gap $\approx$ 1.17 eV and E$_{G}$ is the increased band gap
due to quantum confinement.\\
\par
\hskip 2cm  The value of X is 2 in the usual effective mass approximation.
However, for nanocrystals, the effective mass itself becomes size
dependent and the exponent X has been reported to vary from 1.2 - 1.8 and
becomes 2 at large size$^{( 17, 23 )}$.  For the purpose of our simulation
we consider two cases

\hskip 3cm  a) X = 2, C = 486 eV/\AA$^{2}$ (E$_{GO}$ = 1.71 eV for d = 30
\AA) $^{( 22, 24 )}$ . \\

\hskip 3cm and\\

\hskip 3cm  b) X = 1.4, C = 126 eV/\AA$^{1.4}$ (E$_{GO}$ = 2.25 eV for d =
30 \AA) $^{( 23 )}$ .

\par

\hskip 2cm   The interaction volume of the column structures with incident
light is V(d) = Ld$^{2}$ .  The resulting distribution R(d) = P(d) x V(d)
is properly normalized to have equal number of dipole oscillators in each
case.  We now make a change of variable from d to E$_{G}$ ( using E$_{G}$ =
E$_{g}$ + C/d$^{X}$) to get the corresponding distribution P(E$_{G}$) of
the band gaps.\\
\par
\hskip 2cm  For a Gaussian distribution of sizes (eq.9) - the distribution
of band gaps is \\

$$ P^G(E_G) = P_0(E_G - E_g) ^{(-X-3)/X} exp[- A(({E_{G0}
- E_g\over E_G - Eg})^{1/X}-1)^2] 
\eqno(12)
$$

\noindent where P$_{0}$ is a normalization constant, A =
0.5${(d_{0}/\sigma)}^{2}$ and E$_{b}$ is the maximum value of E$_{G}$.

\par
\hskip 2cm  Following a similar procedure for the Lognormal distribution
(eq.10) we have \\

$$ P^L(E_G) = R_0(E_G - E_g) ^{(-X-2)/X)}\sigma_L^{-1}
exp [-B ({ln(C/E_G - E_g)^{1/X}\over ln(C/E_{G0} - E_g)^{1/X}} -1)^2]
\eqno(13)
$$

\noindent where R$_{0}$ is the normalization constant , B = 0.5
(m$_{0}$/$\sigma_L$)$^{2}$ . \\
\par
\hskip 2cm  Fig. 4 shows the distribution of band gaps P(E$_{G}$) when the
energy upshift is given for X = 1.4 and X = 2.  For the Lognormal case,
we see that P(E$_{G}$) peaks close to the band gap of c-si.\\
\par

\hskip 2cm Equations 8, 12 and 13  are used to calculate $\alpha(E)$ for each
incident energy E = $\hbar\omega$.
\par

\hskip 2cm $\alpha(E)$ for a Gaussian distribution of sizes is given by

$$
\alpha^G(E) = \alpha_0 \int^{E_b}_{0}(E - E_G)^{-W} (E_G - E_g)
^{(\gamma-X-3)/X} exp[- A(({E_{G0} - E_g\over E_G - Eg})^{1/X})-1)^2] dE_G
\eqno(14)
$$

\noindent where $\alpha_0$ is a normalization constant .\\

\par       
\hskip 2cm For a Lognormal distribution of sizes $\alpha(E)$ is 

$$
\alpha^L(E) = \beta_0 \int^{E_b}_{0}(E - E_G)^{-W} (E_G - E_g)
^{((\gamma-X-2)/X)} \sigma_L^{-1} exp[ -B ({ln(C/E_G - E_g)^{1/X}\over
ln(C/E_{G0} - E_g)^{1/X}} -1)^2] dE_G
\eqno(15)
$$

\noindent where $\beta_0$ is a normalization constant. \\
\par
\hskip 2cm  Equations (14) and (15) are used to simulate various possible
absorption processes .

\vskip 2cm

\section{ Results of the simulation }

\hskip 2cm   In this section, we present the results for the absorption
co-efficient as a function of energy for different cases on the basis of
the above model. The values of various exponents for different physical
situations are summarized in Table I.\\
\par
(1) Direct band gap\\
\par
\hskip 2cm   Figure 5 shows $\alpha(\hbar\omega)$ vs $\hbar \omega$ for a
direct band gap material where k is conserved in the optical transition
and no size dependence of the oscillator strengths ( $\gamma$ = 0 ) is
considered (cases 1 and 3 and cases 5 and 7 in Table 1 ). To facilitate
comparison between various cases , we have normalized the absorption to
its peak value in each case. We see from the figure that the absorption
tends to fall off at high energy ( in sharp contrast  with the
experimental data ). These results of the simulation can be easily
understood as follows.
\\
   
\par
\hskip 2cm   For a 1D system, the JDOS has a singularity at  E = $E_{G}$
and falls off for E $>$ E$_{G}$.  The dominant contribution (to each
$\alpha(E)$) is hence only from a particular set of nanostructures with
band gap $E_G$ where E =$ E_{G}$ and all other $E_G$s have almost no
contribution. So the simulated $\alpha(E)$ vs E plot mimics P($E_{G}$).
We hence conclude from the simulation and the experimental data  that
PS is not a direct band gap material and we will not consider this case
further.\\
\par
(2) Indirect band gap\\
\par
\hskip 2cm     We now consider the case when momentum(k) is not conserved
for the optical transition. Here we discuss two cases - size independent
oscillator strengths ( $\gamma$ = 0 ) and size dependent oscillator
strengths ( assuming $\gamma$ = 6 ). \\
\par
\hskip 2cm We first consider the situation where the oscillator strength
has no size dependence. Figure 6 (curves (a) and (c)) is a plot of the
calculated $\alpha$ vs $\hbar\omega$ for a Gaussian distribution of sizes
(case 2 and 4 of table I ). Figure 7 (curves (a) and (c)) is a plot of the
calculated $\alpha$ vs $\hbar\omega$ for a Lognormal distribution of sizes
(case 6 and 8 of Table I). Also shown in the figures is a plot of the
experimental curve for a p type PS. (The dip in $\alpha$ near 2.2 eV in
the experimental curve is an artifact related to a filter change).  To
facilitate comparison , the calculated $\alpha$ is scaled to match the
experimental curve near 3 eV.  We see from the plots 6(a) and 6(c), that
$\alpha$ initially increases with energy and then saturates - at variance
with the experimental results.  We understand this result qualitatively as
follows.  In section I , we have already mentioned that the JDOS is a
constant independent of energy for $\hbar\omega$ $<$ E$_{G}$ and zero for
$\hbar\omega$ $>$ E$_{G}$. Hence for $\hbar\omega$ $>$ $E_G$ ,a given
nanostructure contributes equally at all energies to $\alpha$.  However
P(E$_{G}$) falls off as E$_{G}$ increases. Nanostructures having a band
gap ($E_G$) above a certain cut off in energy (in the high energy tail of
the distribution) do not contribute significantly to $\alpha(\hbar\omega)$
(as P(E$_{G}$) is very small).  This gives rise to the saturation of
$\alpha(E)$ in figures 6(a) and 6(c).This saturation is more obvious in
figures 7(a) and 7(c), because P(E$_{G}$) falls off very rapidly for a
Lognormal distribution of sizes.  These results also do not agree with the
experimental data. \\
\par 
\hskip 2cm The situation is not remedied even if we include higher excited
states (n$_{x}$ = 2, n$_{y}$ = 1; etc.) in the energy range considered in
our calculation.  Based on the above discussion, we see that for $\alpha$
to increase with $\hbar\omega$, it is necessary to increase the
contribution to $\alpha$ from the nanostructures in the high energy region
of $\alpha(E)$ .This is possible if the oscillator strength is enhanced
for the nanostructures with large $E_G$  ( smaller sizes ). \\
\par
\hskip 2cm  We now consider the case when F($E_G$) varies as ${
f/d^\gamma}$.  Figure 6 (curves (b) and (d) for case 2 and 4 in Table I
)and figure 7 (curves (b) and (d) for case 6 and 8 Table I ) are plots of
$\alpha(\hbar\omega)$ vs $\hbar\omega$ for $\gamma$ = 6. For the Gaussian
distribution that we have chosen, we see that the absorption still
saturates  although the onset of saturation moves to slightly higher
energy. This is a measure of the contribution of the (high energy) tail of
$P(E_G)$ for the Gaussian case to the total absorption.  For the Lognormal
distibution, we see from figure 7 (curves (b) and (d)) that
${\alpha(\hbar\omega)}$ vs $\hbar\omega$ now qualitatively looks like the
experimental curve.  \\

\par
\hskip 2cm From a comparison of figures 6 and 7, we find that in order to
get the absorption co-efficient to increase with incident energy , it is
important that the contributions to $\alpha$ from the low energy regions
of P$(E_G)$ do not dominate the optical absorption at large $\hbar\omega$.
This is also satisfied if the distribution of band gaps P(E$_G$) is
sufficiently broad. We illustrate this in Fig.8 where we plot the
calculated value for $\alpha$ for a Gaussian distribution of band gaps
with the mean energy at 2.25 eV and a variance of 0.75 eV.  Hence we find
that the nature of the band gap distribution $P(E_{G})$ plays a crucial
role in determining the shape of the absorption spectrum .\\

\par
\hskip 2cm  We conclude that to properly account for the optical
absorption in PS we need to assume \\
\par
\hskip 2cm (1) PS is a pseudo 1D indirect band gap semiconductor.\\
\par
\hskip 2cm (2) The oscillator strength is size dependent and is given by a
power law ${f/d^{\gamma}}$.\\
\par
\hskip 2cm (3) The distribution $P(E_{G})$ also plays an important role .
Lognormal distribution of the nanostructures diameters in porous silicon
seems to naturally account for the absorption spectrum.\\

\par

\hskip 2cm All these considerations are in general agreement with results
of luminescence experiments on PS .The low temperature luminescence of PS
excited near the peak of the broad luminescence spectrum exhibits steps
which have been associated with the characteristic TO mode of c-silicon
$^{( 25 )}$.  More recently$^{( 20 )}$ well resolved peaks corresponding
to the TO mode of Si have been reported for luminescence of PS suggesting
that PS is an indirect band gap semiconductor.  The observed dominant zero
phonon component of luminescence  constitutes direct evidence for the
enhancement of the oscillator strength due to strong quantum confinement
at smaller sizes . \\
\par
\hskip 2cm  We now comment on the validity of the procedure for obtaining
the band gap of PS from $\sqrt{\alpha\hbar\omega}$ vs $\hbar\omega$.
Figure 9 is a simulated plot of $\sqrt{\alpha\hbar\omega}$ vs
$\hbar\omega$  ( case 8 in Table I ) for three different values of
$\gamma$ and the experimental data on p type sample . First, we note that
the intercept of the linear portion of each of the simulated curves is
neither coincident with the peak position ( near 1.3 eV ) of the
corresponding band gap distribution ( see figure 4 ) nor close to the mean
band gap energy $E_0$ = 2.25  ev. So we see that it is very difficult to
relate this intercept with such physical quantities
.Therefore , it is not obvious how to interpret the intercept with some
kind 'effective band gap '. We  also note that the intercept depends on
the choice of $\gamma$.  We hence believe that the intercept does not have
the simple interpretation of a $band$ $gap$ as defined for a homogeneous
system. We would like to point out that for such a heterogeneous system
the gap can be interpreted only in an operational sense like E$_{03}$ or
E$_{04}$ - depending on the value of the absorption coefficient (after
correcting for the sample porosity).\\
\par
\hskip 2cm Finally we try to understand the temperature dependence of the
band gap. To explain the rigid shift of the absorption curve with
temperature (to lower $\alpha$), we argue that $E_G$ of each of the
nanostructures follows a relation $E_G$ (T)  = $E_{G0} -\gamma T$, where
$E_{G0}$ is the band gap at zero degree Kelvin . We assume that the change
of the band gap with temperature is a consequence, like in bulk
semiconductors, of the electron -phonon interaction and take $\gamma$ =
0.5 meV/K. In our model, changing temperature is equivalent to changing
the value of C in the expression that determines the energy upshift
$\Delta E$.  Figure 10 is a plot of ln$(\alpha)$ vs $\hbar\omega$ for
three different values of C. We see that the absorption curves are
rigidly shifted vertically with respect to each other $^{( 10 )}$ as also
seen in figure 2.  We argue that this provides a $natural$ explanation for
the observed temperature dependence of $\alpha$ . \\

\par

\section{ Conclusions }

\hskip 2cm  In this paper we have investigated and clarified the
absorption process in free standing thin films of PS  . \par

\hskip 2cm  We have elaborately discussed that one can not use eq 1. to
obtain the band gap of porous silicon . We point out that these objections
are also true for other low dimensional nanostructures like quantum wires
, quantum dots etc . One can easily generalize our results ( eq 8 , 14 ,
15 ) and use them to analyse the absorption process for other
nanostructured materials too . For that purpose one just have to put
appropriate density of states function and the size distribution in those
equations .

\hskip 2cm We have calculated the absorption coefiicient as a function of
energy assuming that PS can be modelled as a $pseudo$ 1D system having a
$distribution$ $of$ $band$ $gaps$ . From our calculations, we show that it
is neccessary to look at PS as an $indirect$ $band$ $gap$ material. To
satisfactorily account for the absorption, we need to invoke that the
$oscillator$ $strengths$ $has$ $a$ $size$ $dependence$. We also show that
a Lognormal distribution of diameters of the nanostructures that make up
PS appears to account for the measured absorption spectrum of p PS .
\par

\hskip 2cm We can explain the temperature dependence of the absorption in
PS on the basis of our model too.
\par

\hskip 2cm We have also shown that porosity can be inferred
$non-destructively$ from a transmission measurement in the region of low
absorption .\\

\par
\vskip 3.5cm

\centerline{\bf ACKNOWLEDGEMENT }

\par

\hskip 2cm The authors like to thank Prof. B.M.Arora and Prof. Rustagi for
valuable suggestions .  S.D thanks Sandip Ghosh , Anver Aziz , M.K.Rabinal
and Biswajit Karmakar for useful discussions and A.D.Deorukhkar ,
R.B.Driver , V.M.Upalekar, P.B.Joshi and Subham Majumdar for their help .\\

\newpage
\vskip 4cm
\begin{tabular}{|c|c|c|c|c|cl|r|} \hline\hline
\multicolumn{ 3 }{|c|}{Gaussian distribution in d} & 
  \multicolumn{ 3 }{|c|}{Lognormal distribution in d} \\ \hline 
Case & Value of X     &Nature of Transition &      
Case & Value of X     &Nature of Transition &  \\ \hline 
1.   & 2.0 &  Direct  & 5. & 2.0 &  Direct  &  \\ \hline 
2.   & 2.0 & Indirect & 6. & 2.0 & Indirect &  \\ \hline 
3.   & 1.4 &  Direct  & 7. & 1.4 &  Direct  &  \\ \hline 
4.   & 1.4 & Indirect & 8. & 1.4 & Indirect &  \\ \hline\hline
\end{tabular}  
\vskip 1.5cm

TABLE I. These physical situations we have considered in the model for
the optical absorption in porous silicon .For direct transitions W = 0.5
and for indirect transitions W = 0 . The value of $\gamma$ is taken as 6.0
for size dependent oscillator strengths and zero for the size independent
case.

\newpage

\begin{figure}

\caption{ Porosity calculated  ( using eq. 7 ) from the transmission
curves$^{( 9, 16 )}$ are plotted against the corresponding gravimetrically
determined values of porosity . The points fall nearly on a straight line
of slope one. (a) data from ref. 9 and (b) data from ref. 16 .}
\end{figure}

\begin{figure}
\caption{ Temperature dependence of the absorption spectrum of p type
porous silicon . (a) 300 K and (b) 100 K .}
\end{figure}

\begin{figure}
\caption{ Size distribution of the column widths . (a) For a Gaussian and
(b) for a Lognormal distribution of widths, having d$_{0}$=30 \AA and
$\sigma$= 4 \AA.}
\end{figure}

\begin{figure}
\caption{ Band gap distribution for corresponding size distribution given
in figure 4. (a) and (b) are for Gaussian distribution for X = 2 and 1.4
respectively.(c) and (d) are for the Lognormal distribution for X = 2 and
1.4 respectively. }
\end{figure}

\begin{figure}
\caption { Simulated $\alpha(E)$ vs E plot for $direct$ transitions
($\gamma$ = 0) . Plots (a) and (b) are for the cases (1) and (3) in Table
I .  Plots (c) and (d) are for the cases (5) and (7) in Table I. }
\end{figure}

\begin{figure}
\caption{ Simulated $\alpha(E)$ vs E plot for $indirect$ transitions .(a)
and (b) correspond to case (2) in table I for $\gamma$ = 0 and 6
respectively .(c) and (d) correspond to case (4) in table I for $\gamma$=0
and 6 respectively .(e) Measured $\alpha(E)$ vs E plot for p type PS. }
\end{figure}

\begin{figure}
\caption{ Simulated $\alpha(E)$ vs E plot for $indirect$ transitions .(a)
and (b) correspond to case (6) in table I for $\gamma$=0 and 6
respectively .(c) and (d) correspond to case (8) in table I for $\gamma$=0 and
6 respectively . (e) Measured $\alpha(E)$ vs E plot for p type PS. }
\end{figure}

\begin{figure}
\caption{ (a)Simulated $\alpha(E)$ vs E plot for broad gaussian distribution
of band gaps ( E$_{0}$ = 2.25 eV and $\sigma_E$ = 0.75 eV ) with
$indirect$ transitions , X=1.4 and $\gamma$ = 6 . (b) Measured $\alpha(E)$
vs E plot for p type PS.}
\end{figure}

\begin{figure}
\caption{ Simulated plots of $(\alpha\hbar\omega)^{1/2}$ vs $\hbar\omega$
for (a) $\gamma$ = 5 , (b) $\gamma$ = 5.5 , (c) $\gamma$ = 6.0 and (d)
corresponding plot for the experimental data on p type PS .}
\end{figure}

\begin{figure}
\caption{ Simulated log($\alpha(E)$) vs E plots for various temperatures (
for case 8 with $\gamma$ = 6 ). (a) C = 126 = $C_0$ , (b) C = 1.25$C_0$ ,
(c) C = 1.5$C_0$ . Curves are rigidly shifted vertically from one
another.}
\end{figure}

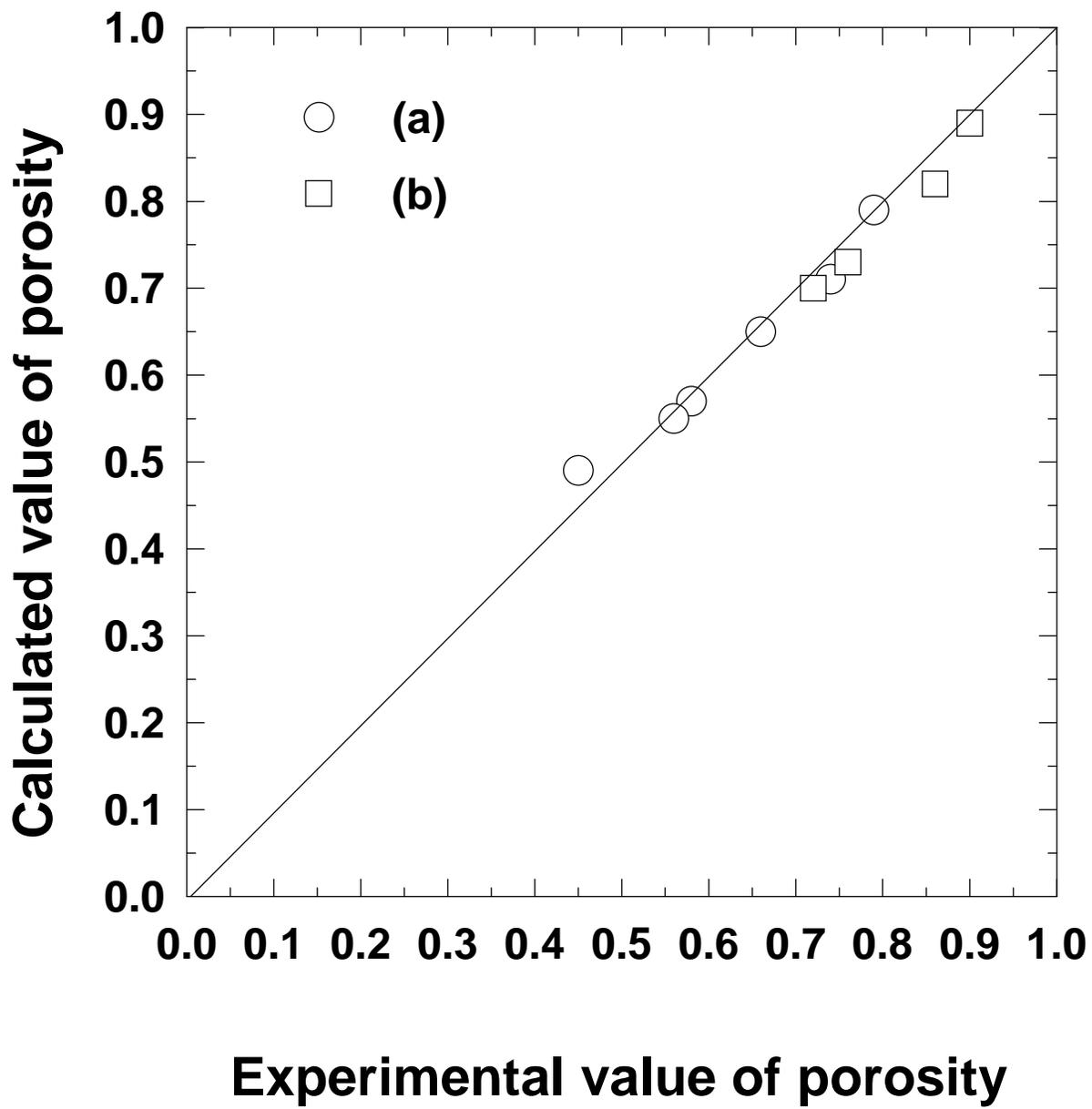
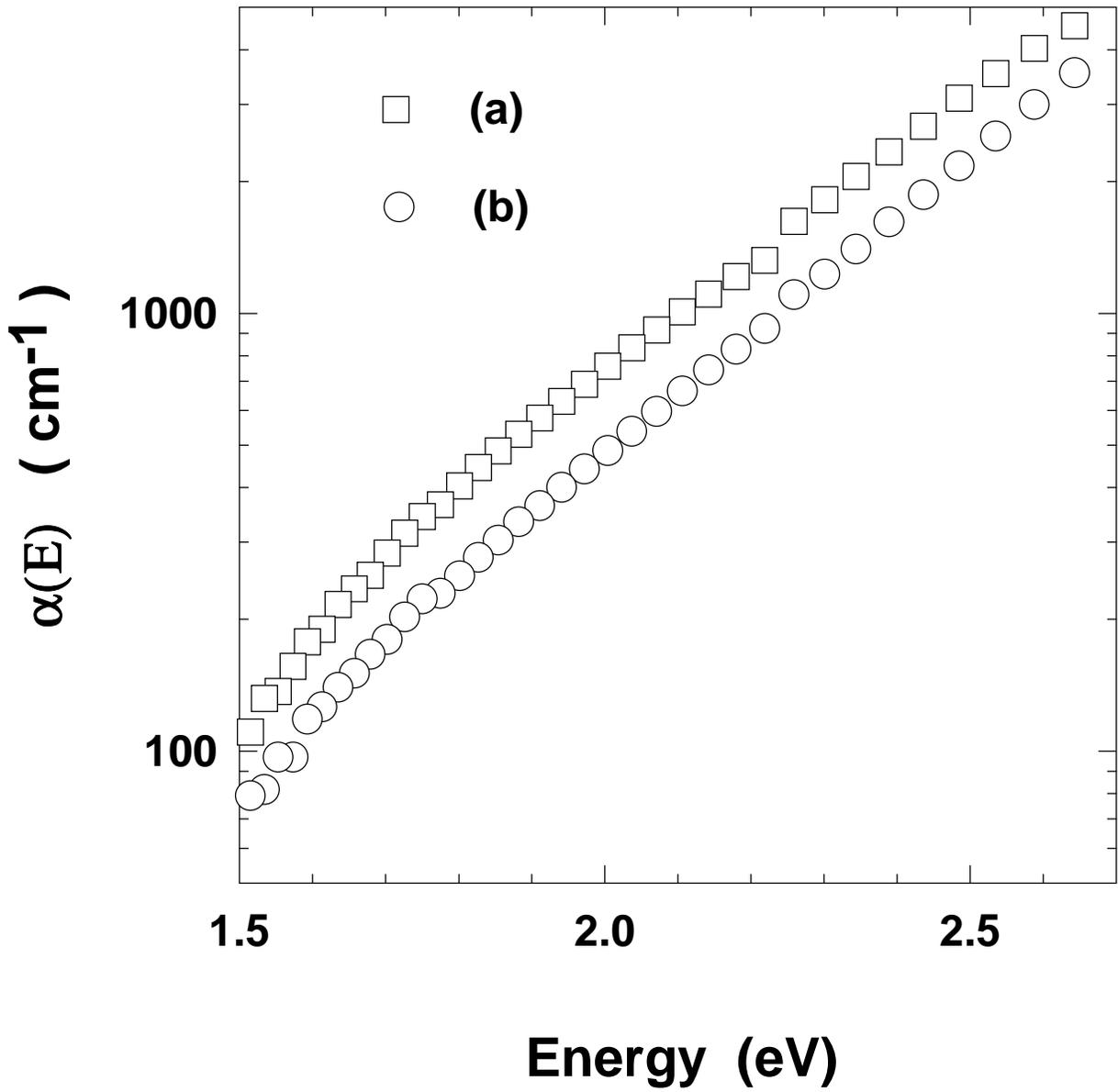
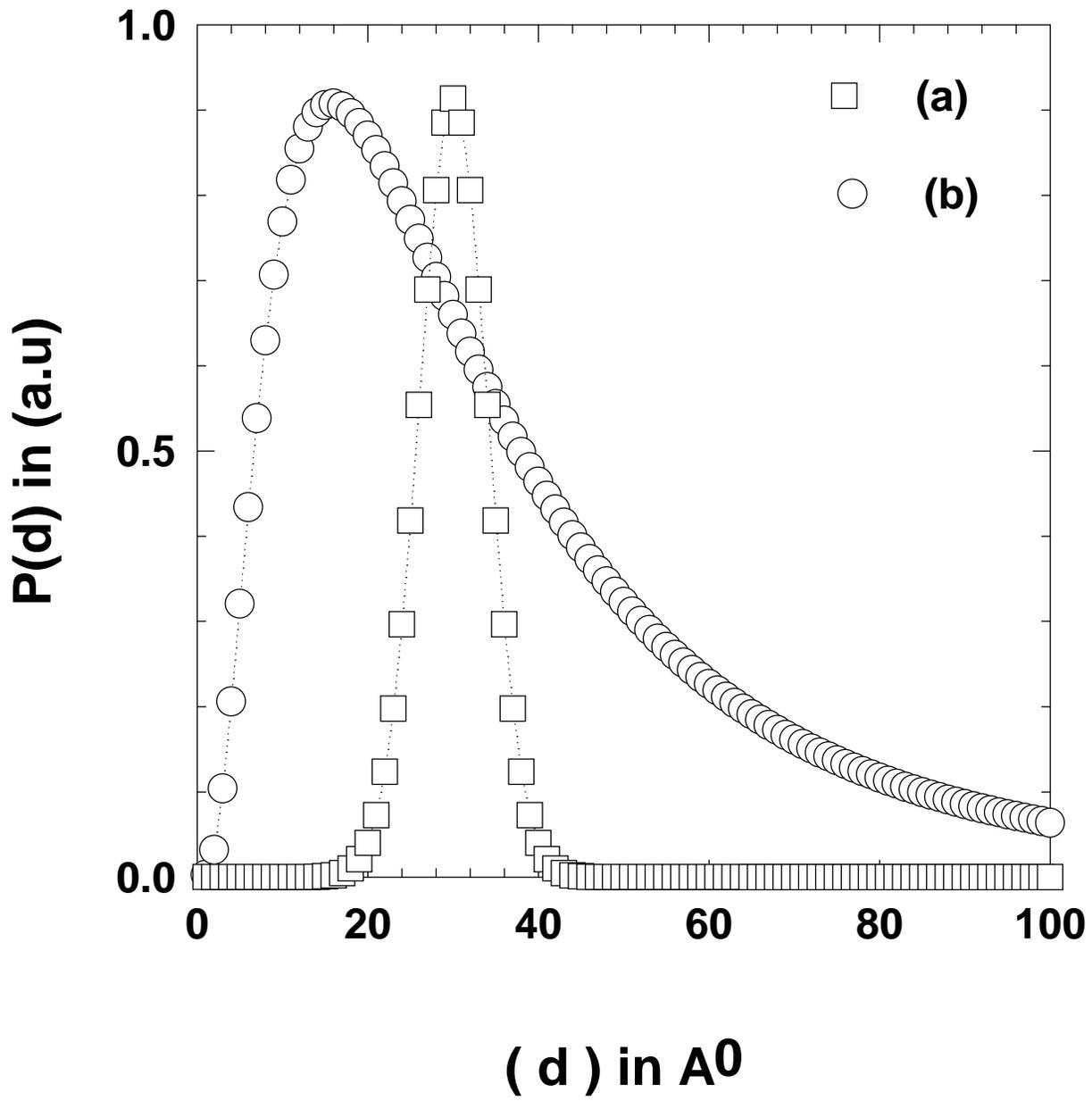
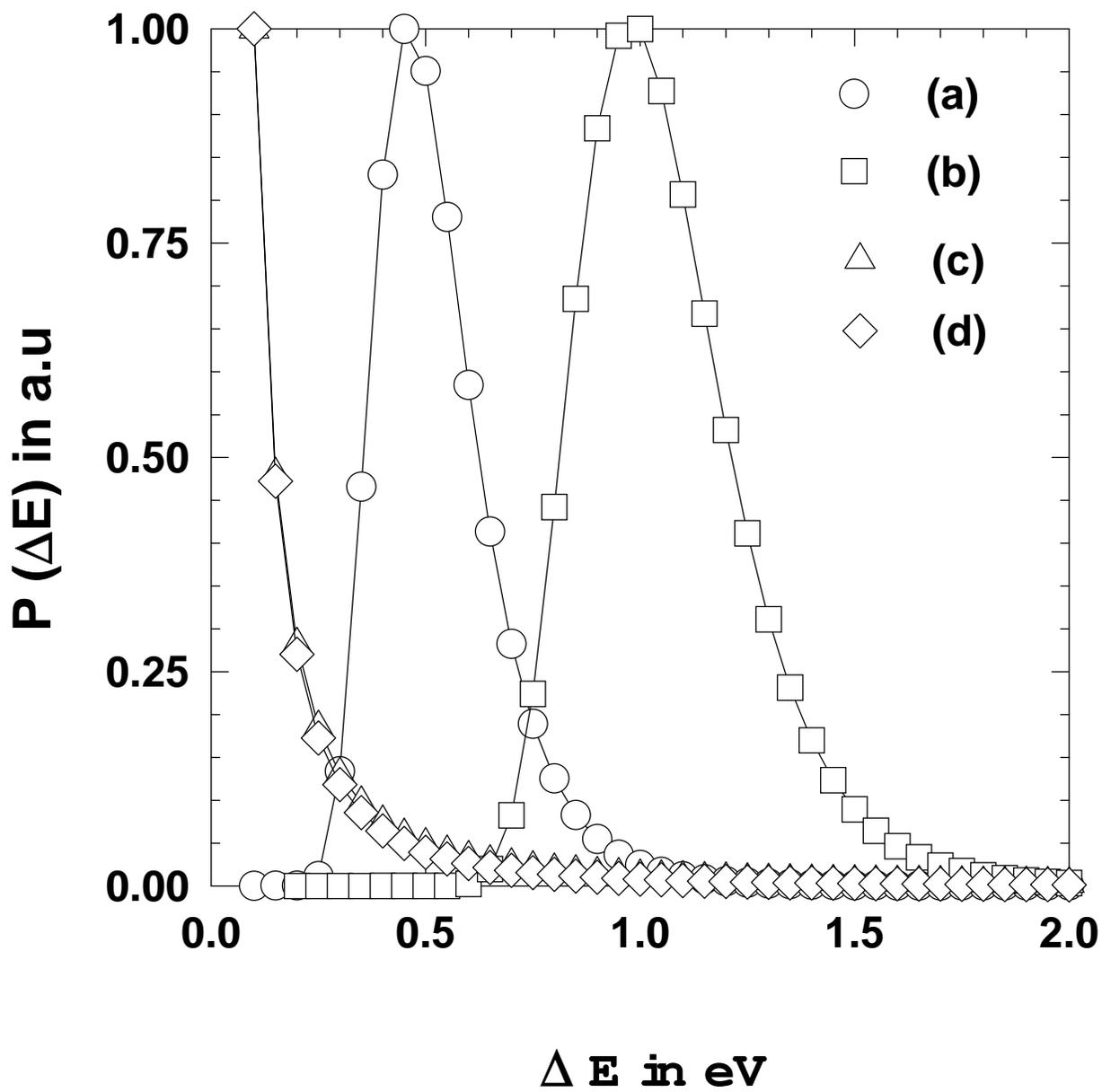
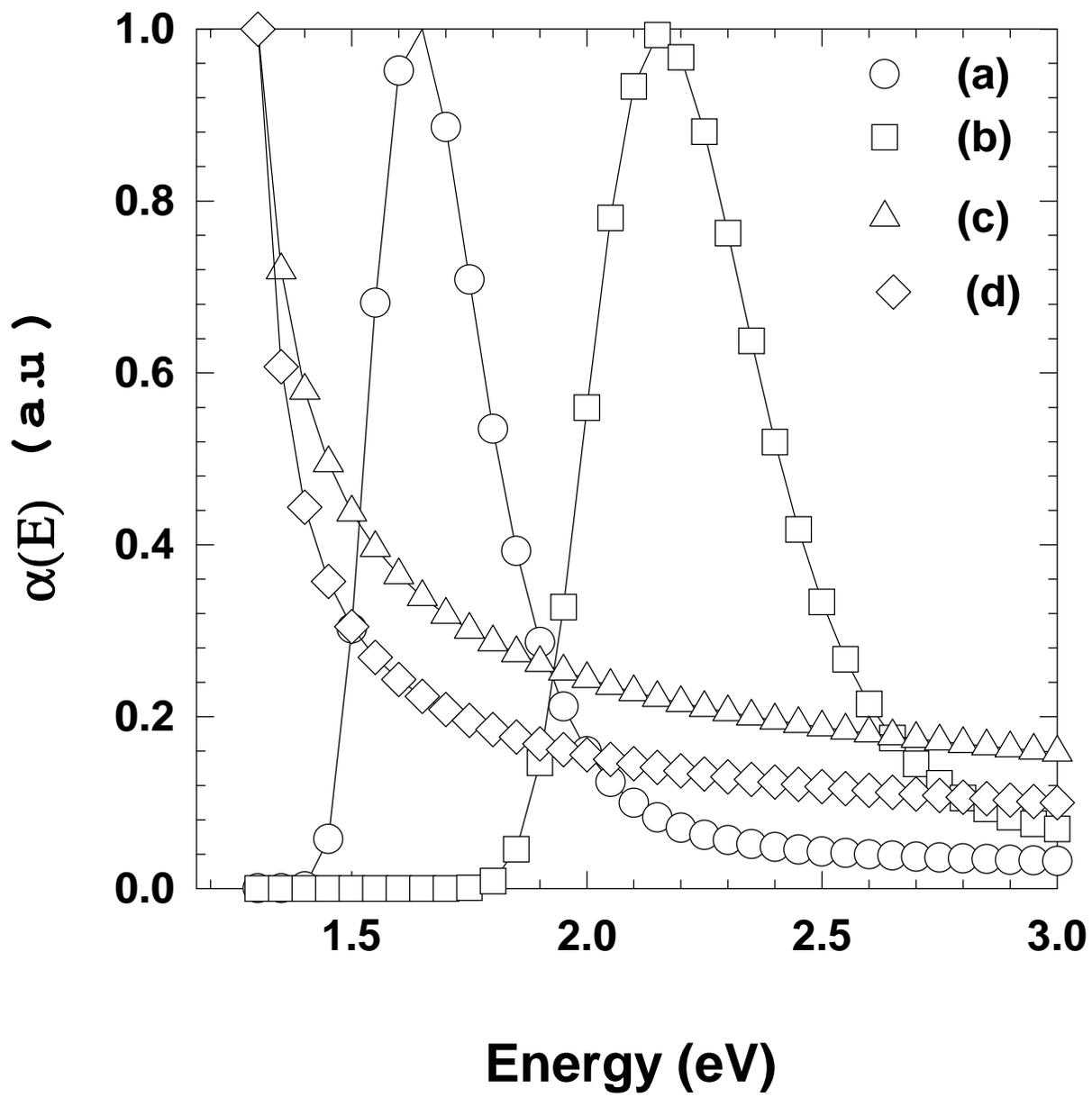
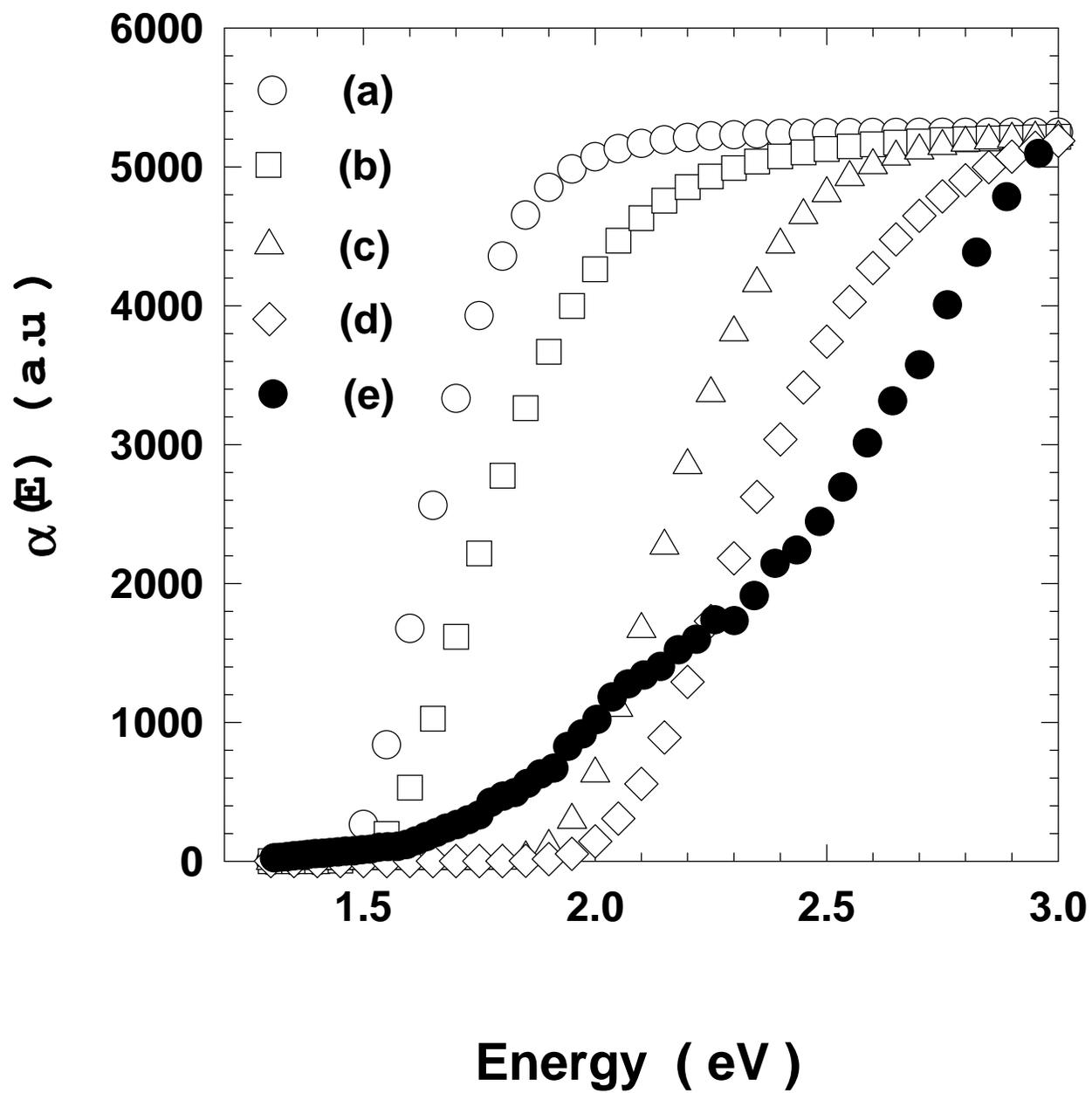
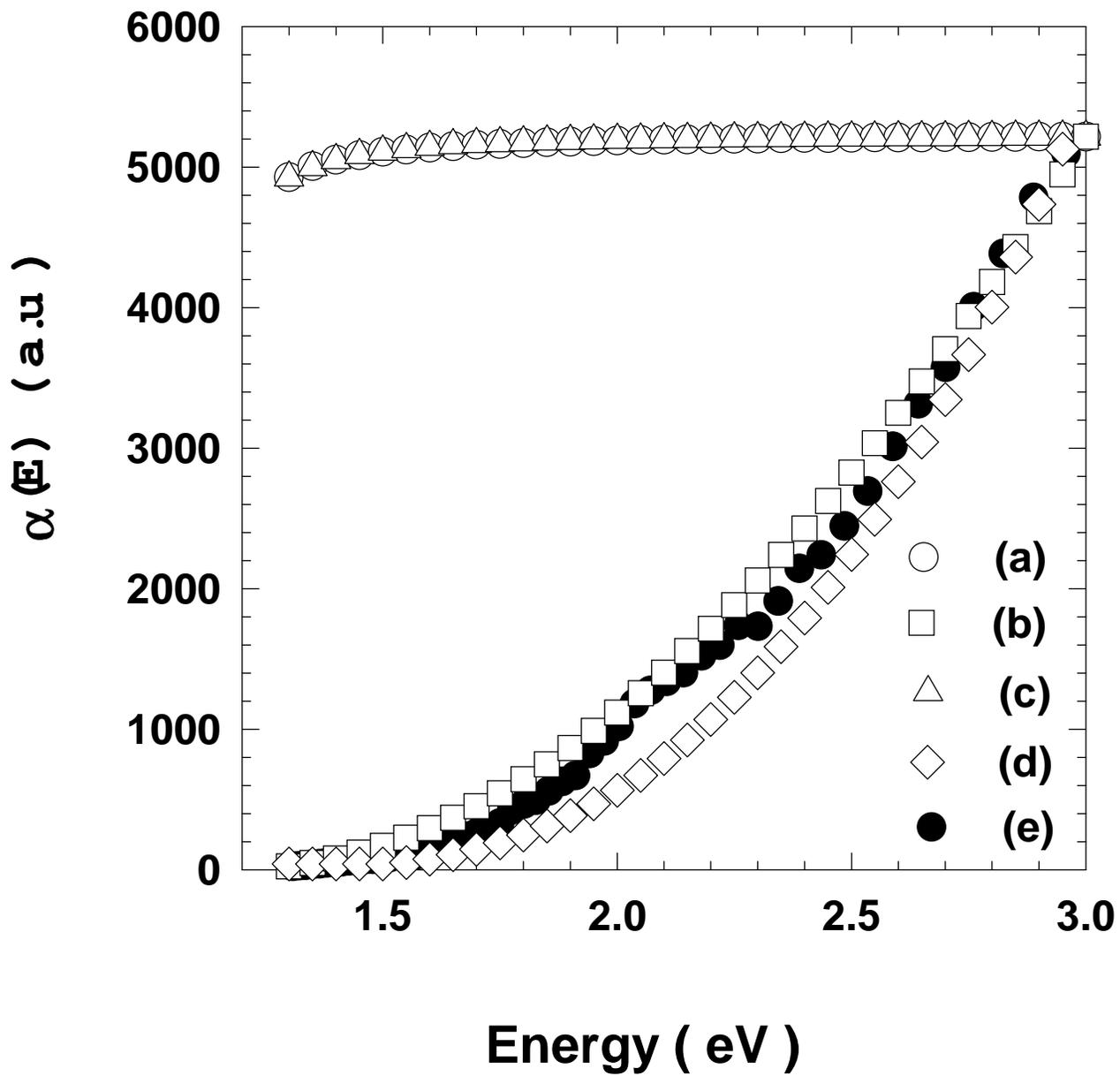
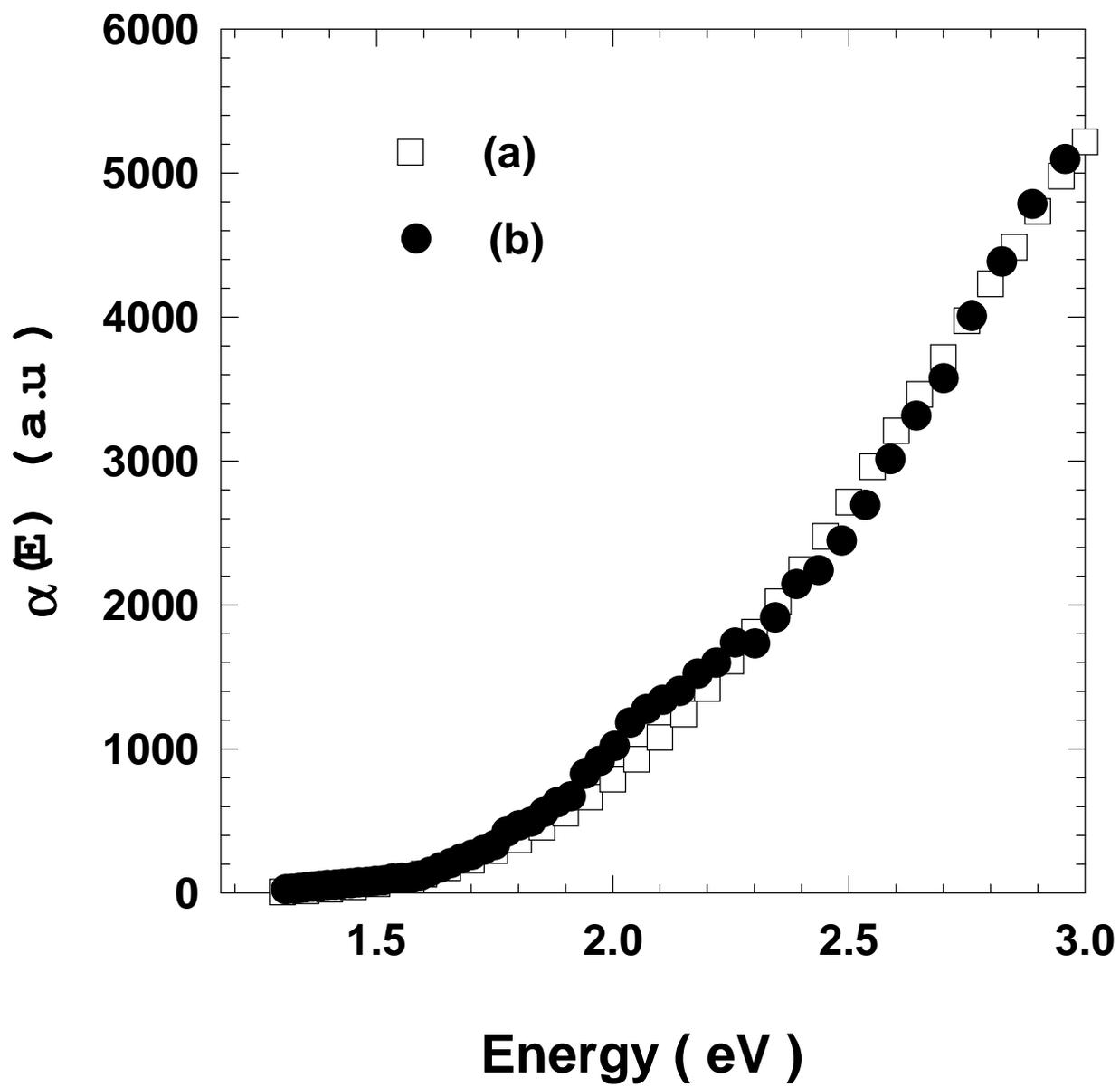
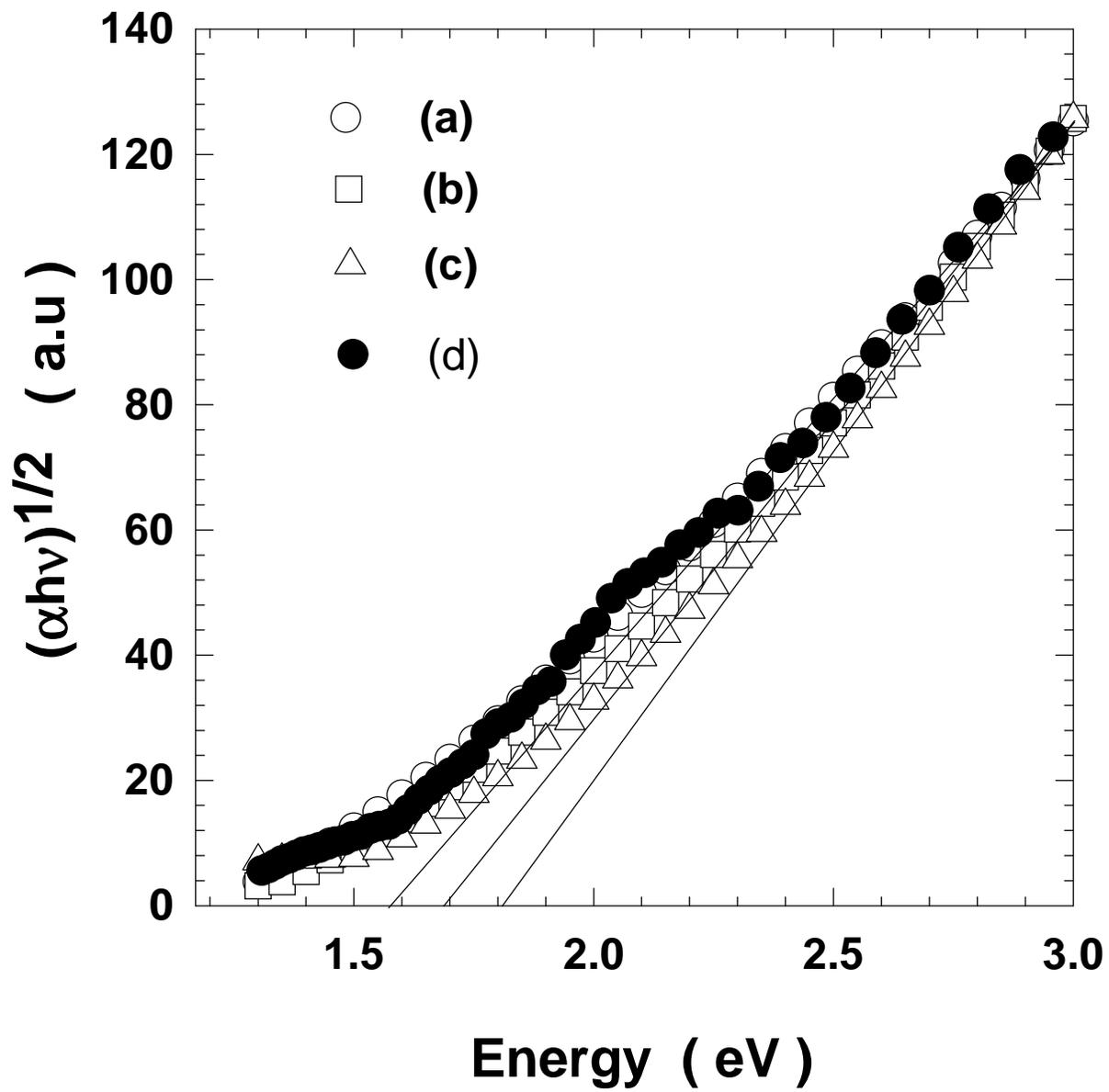
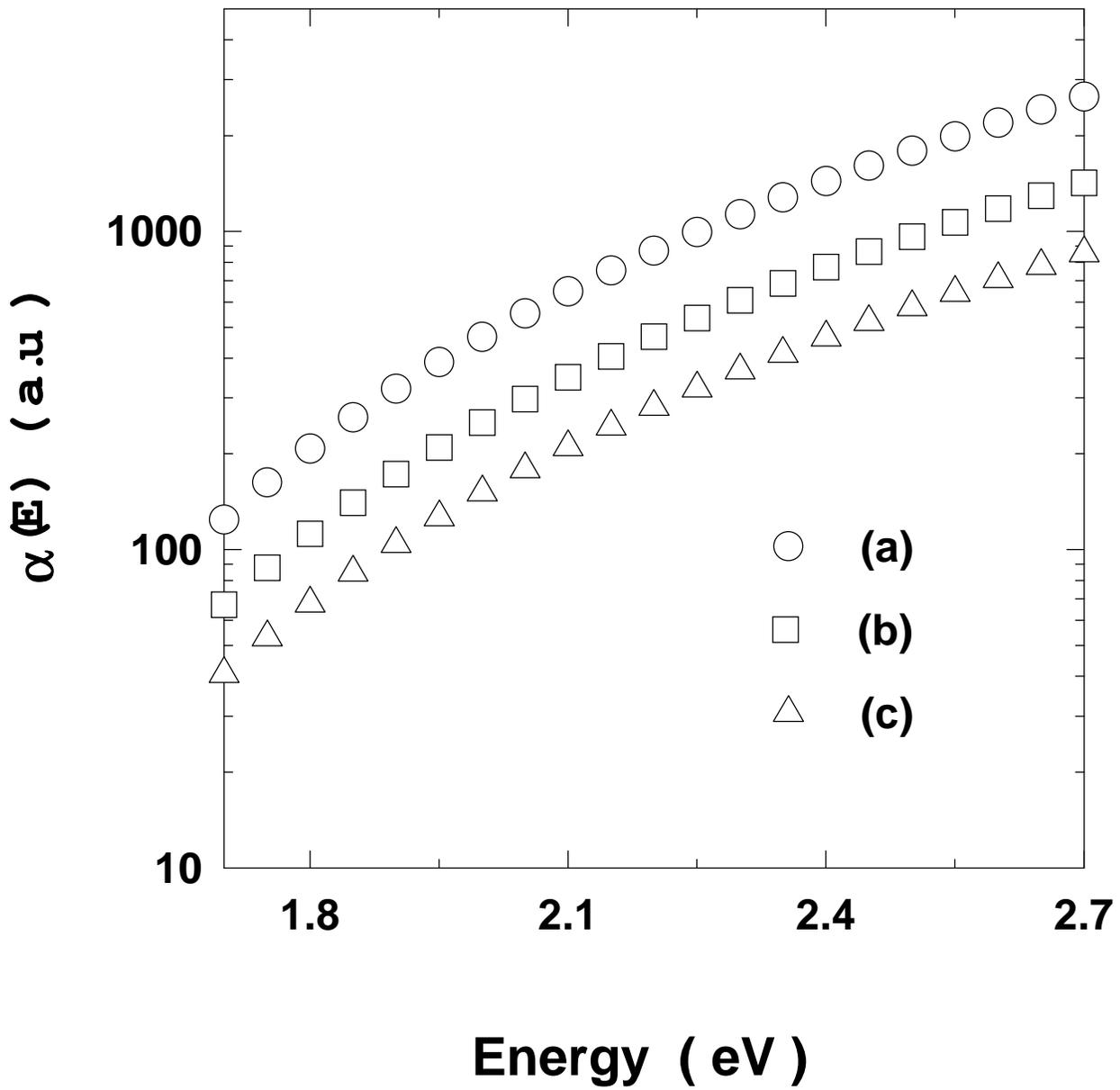
\end{document}